\documentclass[prl,twocolumn,showpacs,amsmath,amssymb]{revtex4}
\usepackage{graphicx}
\usepackage{enumerate}
\newcommand \beq{\begin{eqnarray}}
\newcommand \eeq{\end{eqnarray}}
 
\begin{document}

\title{Bound $H$-dibaryon in Flavor SU(3) Limit of Lattice QCD}

\author{
Takashi Inoue$^{1}$
Noriyoshi Ishii$^{2}$,
Sinya Aoki$^{2,3}$,
Takumi Doi$^{3}$,
Tetsuo Hatsuda$^{4,5}$,\\
Yoichi Ikeda$^{6}$,
Keiko Murano$^{7}$,
Hidekatsu Nemura$^{8}$,
Kenji Sasaki$^{3}$\\ (HAL QCD Collaboration)\\
\bigskip
\includegraphics[width=0.15\textwidth]{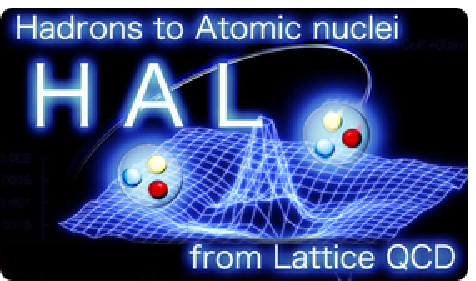}
}

\affiliation{
$^1$Nihon University, College of Bioresource Sciences, Fujisawa 252-0880, Japan\\
$^2$Center for Computational Sciences, University of Tsukuba, Tsukuba 305-8577, Japan\\
$^3$Graduate School of Pure and Applied Sciences, University of Tsukuba,
    Tsukuba 305-8571, Japan\\
$^4$Department of Physics, The University of Tokyo, Tokyo 113-0033, Japan\\
$^5$IPMU, The University of Tokyo, Kashiwa 277-8583, Japan\\
$^6$Nishina Center for Accelerator-Based Science, Institute for Physical \\
    and Chemical Research (RIKEN), Wako 351-0198, Japan\\
$^7$High Energy Accelerator Research Organization (KEK), Tsukuba 305-0801, Japan\\
$^8$Department of Physics, Tohoku University, Sendai 980-8578, Japan
}

\begin{abstract}
 The flavor-singlet $H$-dibaryon, which has strangeness $-$2 and baryon number 2,
 is studied by 
 the approach recently developed for the baryon-baryon interactions in lattice QCD.
 The flavor-singlet central potential is derived from the spatial and
 imaginary-time dependence of the Nambu-Bethe-Salpeter wave function measured in $N_f=3$
 full QCD simulations with the lattice size of $L\simeq 2, 3, 4$ fm.
 The potential is found to be insensitive to the  volume,
 and it leads to a bound $H$-dibaryon with the binding energy of 30--40 MeV
 for the  pseudo-scalar meson mass of 673--1015 MeV.
\end{abstract}
\maketitle

 Search for dibaryons is one of the most challenging theoretical and 
 experimental problems in the physics of strong interaction and quantum chromodynamics (QCD).
 In the non-strange sector, only one dibaryon, the deuteron,
 is known experimentally. In the strange sector, on the other hand, 
 it is still unclear  whether there are bound dibaryons or dibaryon resonances.
 Among others, the flavor-singlet state ($uuddss$), the $H$-dibaryon,
 has been suggested to be the most promising candidate~\cite{Jaffe:1976yi}. 
 The $H$  may also be a doorway to strange matter and to exotic hyper-nuclei~\cite{Farhi:1984qu}.
 Although deeply bound $H$ with the binding energy $B_H > 7 $ MeV from the $\Lambda\Lambda$ threshold
 has been ruled out by the discovery of the double $\Lambda$ nuclei,
 $_{\Lambda \Lambda}^{\ \ 6}$He~\cite{Takahashi:2001nm}, there still  remains a possibility of 
 a shallow bound state or a resonance in this channel~\cite{Yoon:2007aq}.
 
 While  several lattice calculations on $H$ have been reported as reviewed in~\cite{Wetzorke:2002mx}
 (see also recent works \cite{Luo:2007zzb,Inoue:2010hs,Beane:2010hg}),
 there is a  serious problem in studying  dibaryons on the lattice:
 To accommodate two  baryons inside the lattice volume,
 the spatial lattice size $L$ should be large enough. 
 Once  $L$ becomes large, however,  energy levels of two baryons become dense,
 so that quite a large imaginary-time $t$ is required  to make clear
 isolation of the ground state from the excited states.
  All the previous works on dibaryons 
 more or less  face  this issue (see also \cite{Yamazaki:2009ua}).

 The purpose of this Letter is to shed a new light on the $H$-dibaryon 
 by extending the lattice approach recently
 proposed by the present authors \cite{Ishii:2006ec,Inoue:2010hs}.
 Our starting point is the baryon-baryon potential obtained from  the Nambu-Bethe-Salpeter (NBS)
 amplitude measured on the lattice \cite{Ishii:2006ec}.
 Such a potential together with the NBS amplitude can be shown to 
 satisfy the Schr\"{o}dinger type equation and to reproduce the correct phase shifts at low energies.
 It was found on the lattice in the flavor SU(3) limit~\cite{Inoue:2010hs} that, while 
  the celebrated repulsive core of the potential appears in the nucleon-nucleon(NN) channels, 
 the ``attractive core" emerges in the $H$-dibaryon channel.
 These features at the short range part of the potential
 are essentially dictated by the  Pauli exclusion principle in the quark level:
 Six-quarks residing at the same spatial point is partially forbidden
 by the quark Pauli effect in the NN channels, which belong to the 
 flavor 27-plet or 10$^*$-plet, while the
 flavor-singlet six-quarks do not suffer from the Pauli effect \cite{Oka:2000wj} (see also \cite{Kawanai:2010ev}).

 The approach based on the baryon-baryon potential has several advantages. In particular it can be used not only to reduce the finite volume artifact  but also to avoid the problem of contaminations from excited states,  as will be explained later.  
In this Letter, 
 to capture  essential features of the $H$-dibaryon without being disturbed by the quark mass differences, 
 we consider the flavor SU(3) limit where all $u$, $d$, and  $s$ quarks have a common finite mass. 
 This allows us to extract baryon-baryon potentials for irreducible flavor multiplets and 
 to make the comparison among different flavor channels in a transparent manner. 
   
 We start with the NBS wave function \cite{Ishii:2006ec} defined by
 \begin{equation}
 \phi_n(\vec r)
   = \langle 0 \vert (BB)^{(\alpha)}(\vec r,0) \vert W_n; \alpha \rangle,
  \end{equation}
where   the state vector  
 $\vert W_n; \alpha \rangle $ is a QCD eigenstate  with the baryon number 2 (6 quark state)
 and energy $W_n$ in the flavor $\alpha$-plet. 
 $(BB)^{(\alpha)} (\vec r, t)=\sum_{i,j,\vec x} C_{ij}^{(\alpha)} B_i(\vec x+ \vec r,t) B_j (\vec x,t)$
 is a two-baryon operator with a relative distance $\vec r$ in  $\alpha$-plet
  with $B_i$ being a one-baryon composite field operator in the flavor octet. 
 The relation between two-baryon operators in the flavor basis and baryon basis
 are given by the SU(3) Clebsch-Gordan coefficients.

 In the lattice QCD simulations, the above NBS wave functions is extracted from the four point function as
\begin{eqnarray}
\label{eq:G4}
 & & G_4(\vec r, t-t_{0}) 
 = \langle 0|(BB)^{(\alpha)} (\vec r, t)\  \overline{(BB)}^{(\alpha)}(t_0)|0 \rangle 
 \\
 & & = \sum A_n \phi_n(\vec r) e^{- W_n(t-t_0)}, \ A_n=\langle W_n;\alpha \vert  \overline{(BB)}^{(\alpha)}\vert 0\rangle . \nonumber
\end{eqnarray}
 Here $\overline{(BB)}^{(\alpha)}(t_0)$ is a wall source operator at time $t_0$
 to create two-baryon states in  $\alpha$-plet,
 while $(BB)^{(\alpha)} (\vec r,t)$ is the sink operator at time $t$ to annihilate the two-baryon states. 
 Even if  we choose $t-t_0$ moderately large so that the  inelastic scatterings
 (e.g. the scattering with excited baryons and the scattering with meson production)
 do not  contribute to $G_4$, there still remain elastic scattering states with low energy excitations
 due to the relative motion of the baryons. 
 For example, with the baryon mass $M \simeq$ 2 GeV in a finite box of $L=4$ fm,
 the non-interacting two-baryon system has  $W_1-W_0 \simeq (2\pi/L)^2/(2 \mu) \simeq$ 50 MeV, 
 with the reduced mass $\mu=M/2$.
 This requires $t-t_0 >$ 10 fm to achieve  1/10
 suppression of the first excited state $\phi_1(\vec r)$ in  $G_4(\vec r, t-t_{0})$.
 It is beyond most of the previous and current lattice simulations.
 
 Our potential approach avoids the above
   problem in the following way:
 The two-body potential in low energy QCD
 dictates all the elastic scattering states 
 $\phi_n (\vec r,t) = \phi_n(\vec r) e^{- (W_n-2M) t}$ simultaneously  
  through the Schr\"{o}dinger equation in the Euclidean space-time \cite{Ishii:2006ec}.
  With the non-relativistic approximation for $W_n$, it reads
 \begin{eqnarray}
 \label{eq:SEn}
 H_0 \phi_n({\vec r}, t) + \int d^3r' U(\vec r, {\vec r}')\phi_n({\vec r}', t)
 = - \frac{\partial}{\partial t} \phi_n({\vec r}, t),    
 \end{eqnarray}
 where $H_0=-\nabla^2/(2\mu)$ and $U$ is a non-local and energy-independent potential. 
 Since the above equation  is linear in $\phi_n$, the linear combination
 such as $\phi(\vec r,t) \equiv \sum_n A_n \phi_n(\vec r,t)=G_4(\vec r,t)/e^{-2Mt}$ also satisfies Eq.(\ref{eq:SEn}).
 We note  that the derivative expansion of $U$ in terms of its non-locality leads to
 $U({\vec r}, {\vec r}')= [V_C(r) + V_T (r) S_{12} + V_{LS} (r) {\vec L}\cdot {\vec S}
 + \cdots )   \delta ({\vec r} - {\vec r}')$ \cite{Ishii:2006ec},
 where $V_C, V_T$ and $V_{LS}$ are the central, tensor and spin-orbit potentials, respectively,
 and dots stands for terms including power of $\nabla$.
 It was shown in \cite{Murano:2010hh} that the leading order potentials without $\nabla$ 
 dominate the potential  at low energies. 
 Thus, the relevant term in the spin-singlet channel, $V_C$, is obtained as 
\begin{eqnarray}
  V_C(r) = \frac{(-H_0 -\frac{\partial}{\partial t} )\phi(\vec r,t)}{\phi(\vec r,t)}.
\end{eqnarray}
 In this way, one can extract the baryon-baryon potential 
 without identifying each elastic states $\phi_n(\vec r,t)$ 
 as long as  $t-t_0$ is so chosen that the inelastic scatterings are suppressed.
 Once we obtain the volume independent $V_{C}$,  binding energies and  scattering phase shifts in the infinite volume
 are obtained by solving the Schr\"{o}dinger equation.
 In contrast to the conventional L\"{u}scher's method \cite{Luscher:1990ux},
 we do not calculate the energy shift of two hadrons at finite $L$
 to access the observables at $L \rightarrow \infty$.
 Further theoretical details of this method will be given in a separate publication \cite{HAL_new}.

 Let us now consider the interaction between flavor-octet baryons in the flavor SU(3) limit, 
 for which two baryon states with a given angular momentum
 are labeled by the irreducible flavor multiplets as
 ${\bf 8} \otimes {\bf 8}=({\bf 27} \oplus {\bf 8}_s \oplus {\bf 1})_{\rm symmetric} 
  \oplus ({\bf 10}^* \oplus {\bf 10} \oplus {\bf 8}_a )_{\rm anti-symmetric}$.
 Here ``symmetric" and ``anti-symmetric" stand for the symmetry under the
 flavor exchange of two baryons.
 For the system in the orbital S-wave, the Pauli principle between two baryons imposes 
 ${\bf 27}$, ${\bf 8}_s$ and ${\bf 1}$ to be spin singlet  ($^1S_0$) while 
 ${\bf 10}^*$, ${\bf 10}$ and ${\bf 8}_a$ to be spin triplet ($^3S_1$). 
 Since  different multiplets are independent  in the flavor SU(3) limit, 
 one can define the corresponding potentials as
 $V^{({\bf 27})}(r), \ V^{({\bf 8}_s)}(r), \ V^{({\bf 1})}(r)$ for $^1S_0$   
 and $ V^{({\bf 10}^*)}(r), \ V^{({\bf 10})}(r), \ V^{({\bf 8}_a)}(r) $ for $^3S_1$.
 Hereafter, we focus  on  the flavor-singlet channel with
\begin{eqnarray}
 BB^{(1)} = - \sqrt{\frac{1}{8}}\Lambda\Lambda
            + \sqrt{\frac{3}{8}}\Sigma\Sigma
            + \sqrt{\frac{4}{8}}N\Xi ,
\end{eqnarray}
 where $\Lambda$, $\Sigma$, $N$ and $\Xi$ are the standard baryon operators
 with Lorentz structure, $[q(C\gamma_5)q]q$~\cite{Inoue:2010hs}.
 
\begin{table}[t]
\caption{\label{tbl:lattice} 
 Summary of lattice parameters and hadron masses. 
 The uncertainty of  $a$~\cite{CPPACS-JLQCD} is not reflected in hadron masses.
 }
  \begin{tabular}{c|c|c|c|c|c}
   \hline \hline
  $a$ [fm] & $L$ [fm] &  $\kappa_{uds}$  & ~$m_{\rm ps}$ [MeV]~ & ~ $m_{B}$ [MeV]~ & ~$N_{\mbox{cfg}}$~ \\
   \hline 
              &      &   ~0.13710~ & 1015.0(6) & 2030(2) & 360 \\
   {0.121(2)} & 3.87 &   ~0.13760~ & ~836.5(5) & 1748(1) & 480 \\
              &      &   ~0.13800~ & ~672.9(7) & 1485(2) & 240 \\
   \hline \hline
 \end{tabular}
\end{table}

 In our dynamical lattice QCD simulations, 
 we employ the renormalization group improved Iwasaki gauge action
 and the non-perturbatively $O(a)$ improved Wilson quark action.  
 For $16^3 \times 32$ lattice, we use the configuration set 
 generated by CP-PACS and JLQCD Collaborations \cite{Ishikawa:2007nn} at $\beta = 1.83$.
 In addition, we generate gauge configurations with the same $\beta$
 for $24^3 \times 32$ and $32^3 \times 32$ lattices, using the DDHMC/PHMC code~\cite{Aoki:2008sm}.
 Quark propagators are calculated  for the spatial wall source at $t_0$
 with the Dirichlet boundary condition in the temporal direction.
 The sink operator is projected to the $A_{1}^{+}$ representation of the cubic group,
 so that the NBS wave function is dominated by the S-wave component.
 For the time derivative, we adopt the symmetric difference on the lattice. 
 Lattice parameters such as lattice spacing $a$, the hopping parameter $\kappa_{uds}$,
 the number of configurations $N_{\rm cfg}$,  together with
 the pseudo-scalar meson mass $m_{\rm ps}$ and the octet baryon mass $m_B$ are summarized
 in Table \ref{tbl:lattice}  for $32^3 \times 32$  lattice.

\begin{figure}[t]
\begin{center}
\includegraphics[width=6.5cm]{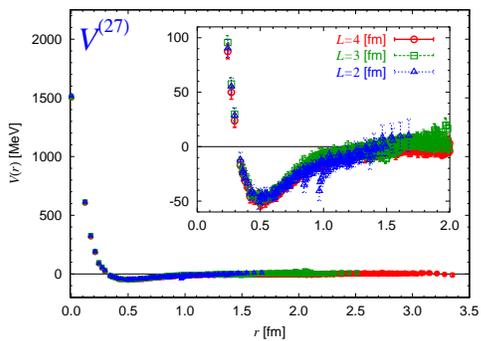}
\end{center}
\vspace{-0.5cm}
\caption{Flavor 27-plet potential $V_C^{(27)}(r)$ obtained for
lattice sizes $L=1.94,\, 2.90,\, 3.87$ fm
at $m_{\rm ps}=1015$ MeV and 
$(t-t_0)/a = 10$.
}
\label{Fig1}
\end{figure}

 To check the qualitative consistency with previous works, 
 we show in Fig.\ref{Fig1} the central potential in the 27-plet channel $V_C^{(27)}(r)$
 obtained in three different lattice volumes with $L=1.94,\,2.90,\,3.87$ fm
 at $m_{\rm ps}=1015$ MeV and $(t-t_0)/a=10$.
 This is the case corresponding to the NN potential in the $^1S_0$ channel.
 Compared with statistical errors, the $L$ dependence is found to be negligible. 
 The $t$ dependence is also small as long as $(t-t_0)/a \geq 9$. 
Note that we do not need overall shift of the potential:  it approaches
 zero automatically as $r$ increases.  The figure shows a repulsive core at short distance 
 surrounded by an attractive well at medium and long distances, which is
  qualitatively consistent with our 
 previous results in quenched and  full QCD simulations reviewed in \cite{Ishii:2010th}.

\begin{figure}[t]
\begin{center}
\includegraphics[width=6.5cm]{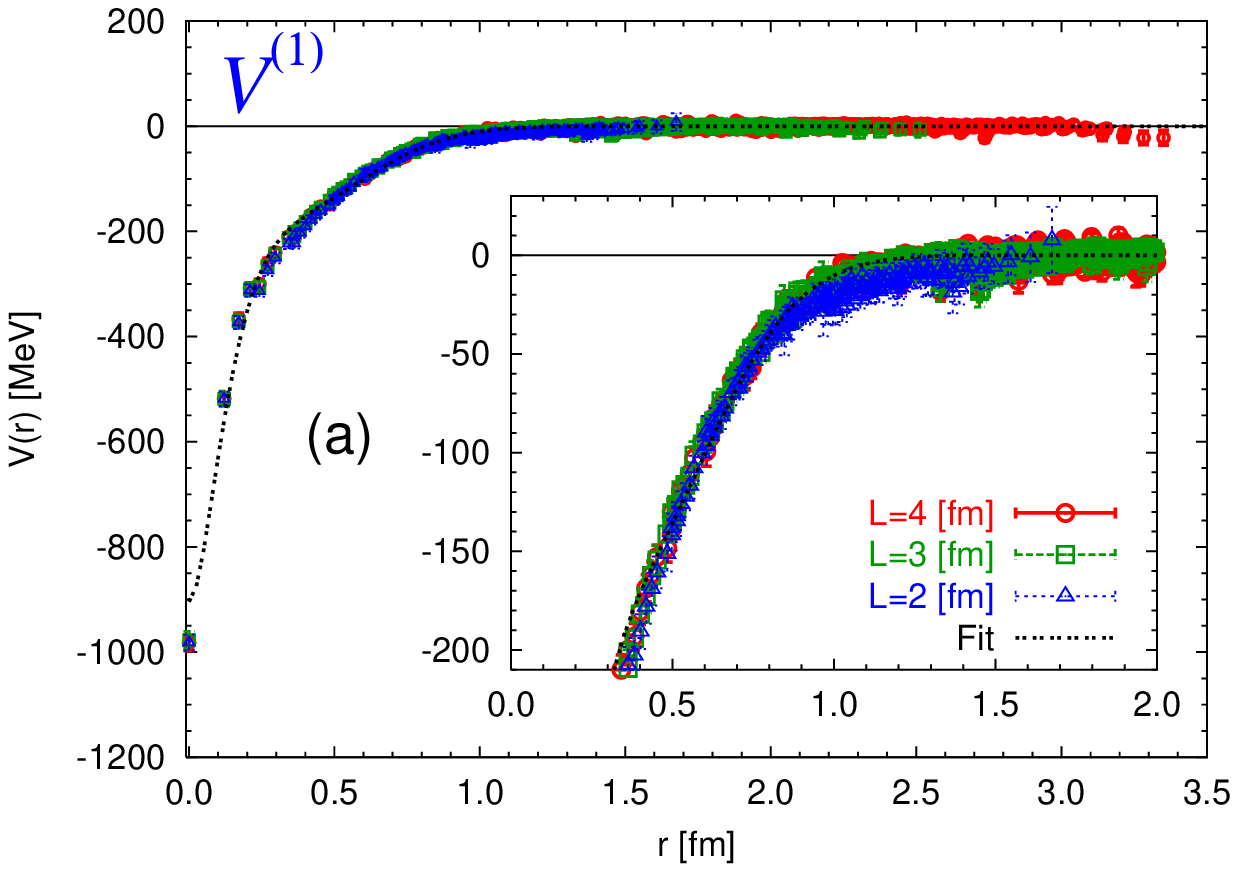}
\vspace{0.3cm}
\includegraphics[width=6.5cm]{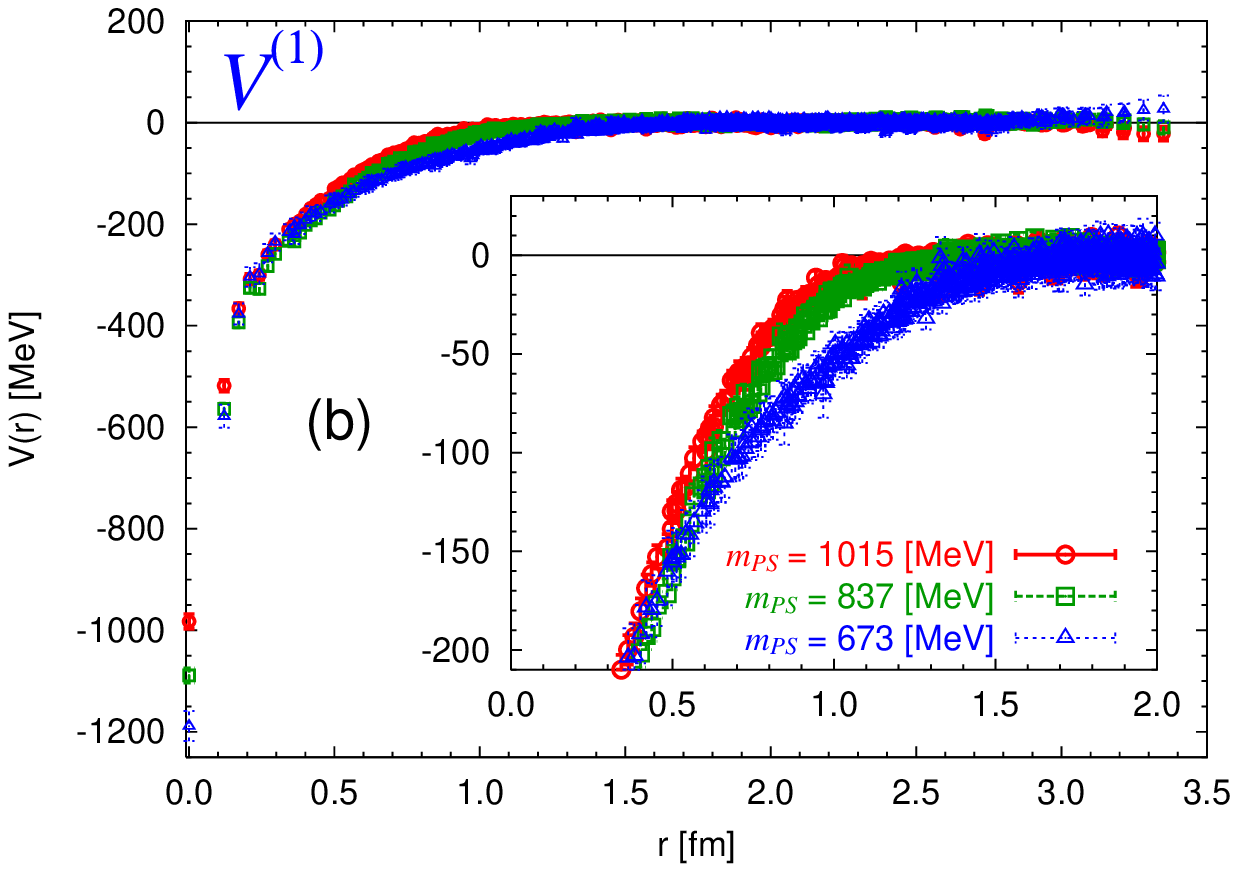}
\end{center}
\vspace{-0.5cm}
\caption{ Flavor-singlet potential $V_C^{(1)}(r)$ at $(t-t_0)/a = 10$. 
 (a) Results for $L=1.94,\, 2.90,\, 3.87$ fm at $m_{\rm ps}=1015$ MeV.
 (b) Results for $L=3.87$ fm at  $m_{\rm ps}=1015,\, 837,\, 673$ MeV. 
  }
\label{Fig2}
\end{figure}

 Shown in Fig.\ref{Fig2}(a) and  Fig.\ref{Fig2}(b)
 are the volume dependence and the quark mass dependence 
 of the central potential in the flavor-singlet channel $V_C^{(1)}(r)$, respectively.
 In both figures, we take $(t-t_0)/a=10$ and have checked that the potentials
 do not have appreciable change with respect to the  choice of $t$.
 We find that the flavor-singlet potential has an ``attractive core" and its range is well localized in space.
 Because of the latter property, we find no significant volume dependence of the potential within the statistical 
 errors as seen in Fig.\ref{Fig2}(a).  
 We find that the long range part of the attraction tends to increase 
 as  the quark mass decreases [Fig.\ref{Fig2}(b)].

 We fit the resultant potential by the following analytic function composed of
 an attractive Gaussian core plus a long range (Yukawa)$^2$ attraction:
$
  V(r) = b_1 e^{-b_2\,r^2} + b_3(1 - e^{-b_4\,r^2})\left( {e^{-b_5\,r}}/{r} \right)^2 .
$
 With the five parameters, $b_1$ -- $b_5$,  we can fit the function to the lattice results
 reasonably well with $\chi^2/{\rm dof} \simeq 1$.
 The fitted result for $L=3.87$ fm is shown by the dashed line in Fig.\ref{Fig2}(a).
 
\begin{figure}[t]
\begin{center}
\includegraphics[width=6.5cm]{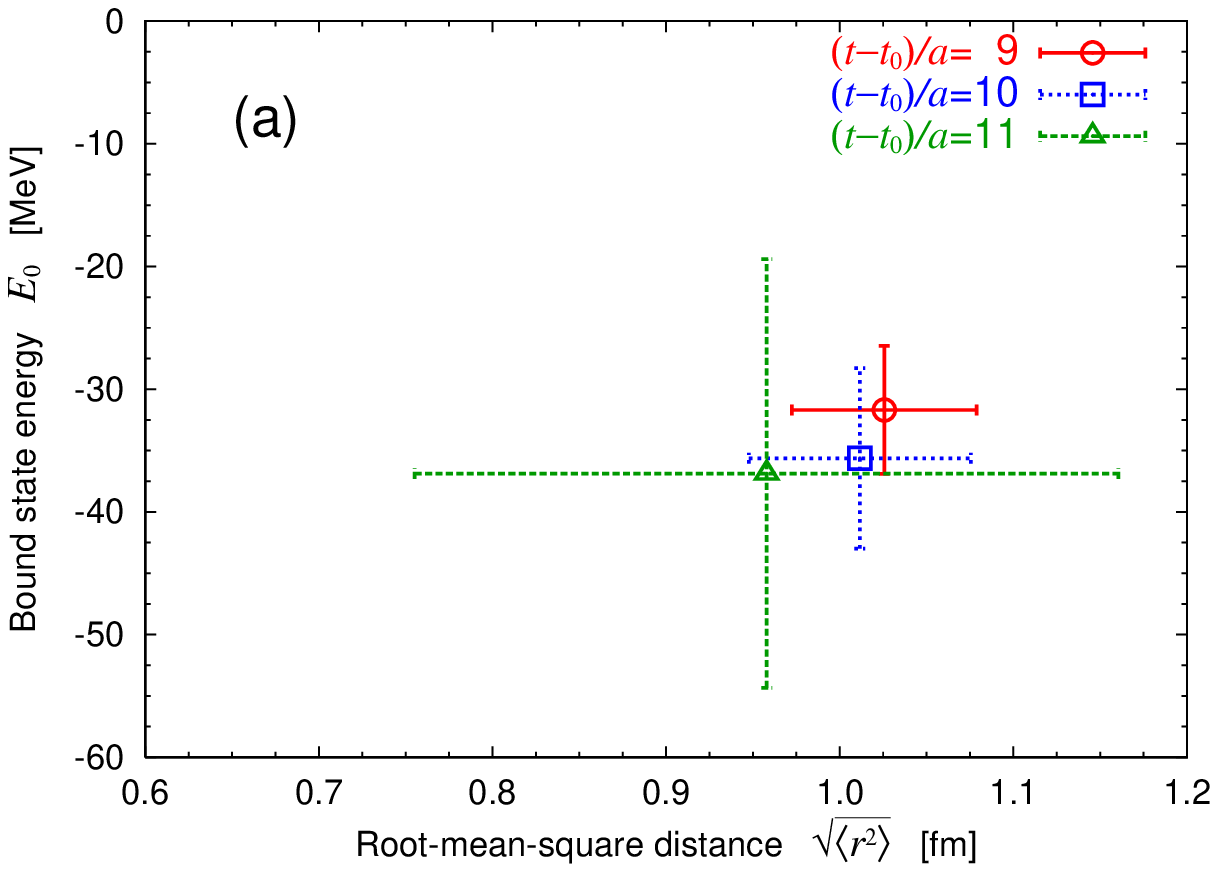}
\vspace{0.3cm}
\includegraphics[width=6.5cm]{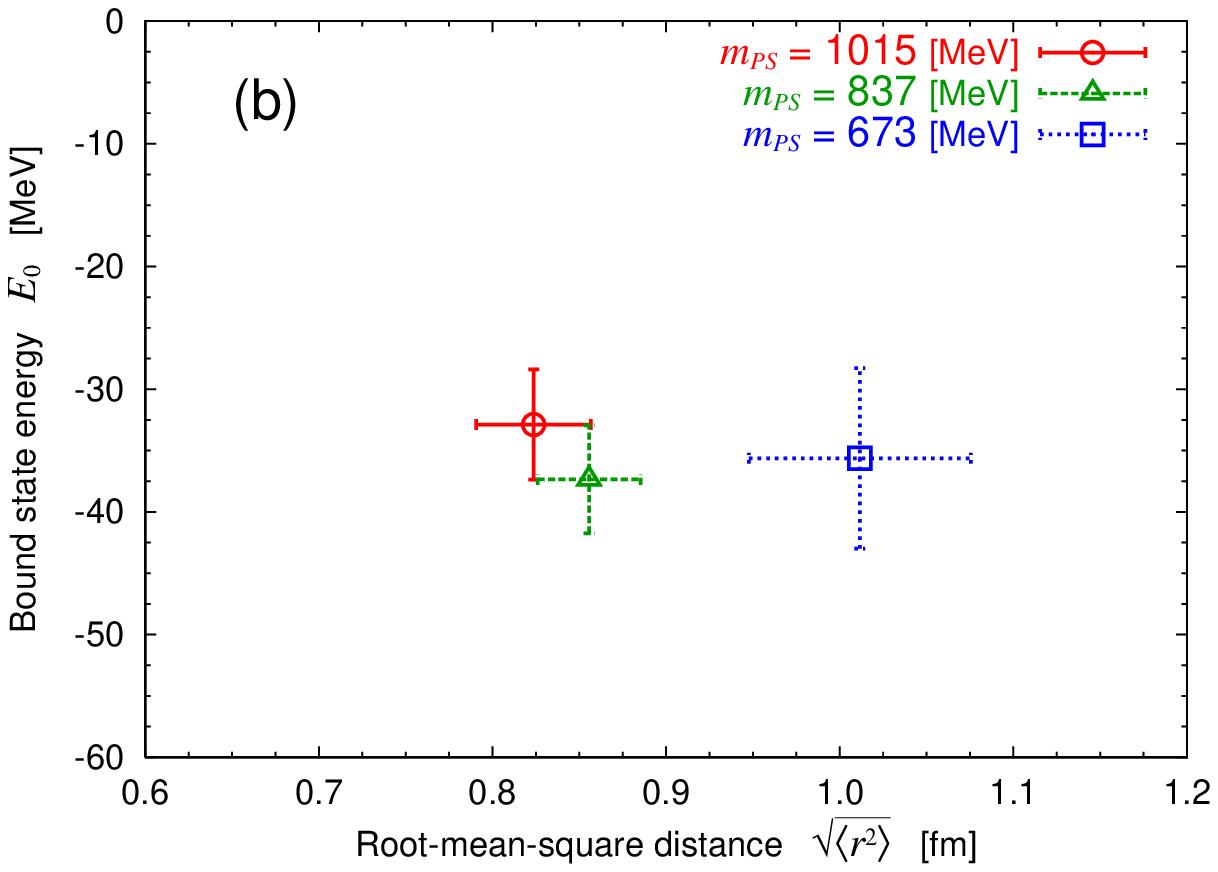}
\end{center}
\vspace{-0.5cm}
\caption{
Bound state energy $E_0\equiv -{\tilde B}_H$ and the rms distance $\sqrt{\langle r^2\rangle}$ of the $H$-dibaryon 
obtained from the potential at $L=3.87$ fm.
(a) Imaginary-time dependence at $m_{\rm ps}=673$ MeV. 
(b) Quark mass dependence at $(t-t_0)/a=10$.}
\label{Fig3}
\end{figure}

 Finally, using the potential fitted by the function,
 we solve the Schr\"{o}dinger equation in the infinite volume
 and obtain the energies and the wave functions for the present quark masses in the flavor SU(3) limit.
 It turns out that, in each quark mass, there is only one bound state
 with the binding energy of  30--40 MeV.  
 In Fig.\ref{Fig3}(a), 
 the energy and the root-mean-square (rms) distance 
 of the bound state are plotted in the case of 
 $(t-t_0)/a=9, 10, 11$ at $m_{\rm ps}=673$ MeV and  $L=3.87$ fm, where
 errors are estimated by the jackknife method.  
 Although the statistical error increases as $t$ increases,
 we observe small changes of central values, which will be included as the systematic errors in our final results.
 Fig.\ref{Fig3}(b) shows the energy and the rms distance of the bound state at each quark mass 
 obtained from the potential with  $L=3.87$ fm and $(t-t_0)/a=10$. 
 Despite that the potential has quark mass dependence,
 the resultant binding energies of the $H$-dibaryon are insensitive in the present range of the quark masses.
 This is due to the fact that the increase of the  attraction toward the lighter quark
 mass is partially compensated by the increase of the kinetic energy for the lighter baryon mass. 
 It is noted that there appears no bound state for the potential of the 27-plet channel 
 in the present range of the quark masses.
   
The final results of the binding energy in the SU(3) limit ${\tilde B}_H$
and the rms distance $\sqrt{\langle r^2\rangle}$ are given below,
with  statistical errors (first) and systematic errors from the $t$-dependence(second).
\begin{eqnarray}
 m_{ps}=1015 ~\mbox{MeV} :~ \  {\tilde B}_H &=& 32.9 (4.5)(6.6) ~\mbox{MeV} \nonumber \\
~~  \sqrt{\langle r^2 \rangle} &=& 0.823(33)(40)  ~\mbox{fm} \nonumber \\
 m_{ps}=~837 ~\mbox{MeV} :~ \  {\tilde B}_H &=& 37.4 (4.4)(7.3) ~\mbox{MeV} \nonumber \\
~~  \sqrt{\langle r^2 \rangle} &=& 0.855(29)(61)  ~\mbox{fm}  \nonumber \\
 m_{ps}=~673 ~\mbox{MeV} :~ \  {\tilde B}_H &=& 35.6 (7.4)(4.0) ~\mbox{MeV} \nonumber \\ 
~~  \sqrt{\langle r^2 \rangle} &=& 1.011(63)(68)  ~\mbox{fm}
 \nonumber
\end{eqnarray} 
A less than 1\%  error from the choice for the fit function is not included here.

Since $\tilde{B}_H$  has the weak quark mass dependence,
one may assume a similar binding energy is realized 
even with the realistic SU(3) breaking, where
$\tilde{B}_H$ is interpreted as the binding energy from the average mass of  two octet baryons in the $S=-2$ and $I=0$ channel. 
Considering that the difference between this average and  $2m_\Lambda$ is about the same amount to $\tilde{B}_H$, 
the $H$-dibaryon may appear as a weakly bound state  or a resonant state near the $\Lambda\Lambda$ threshold, 
as mentioned in \cite{Inoue:2010hs}.
To make a definite conclusion on this point, however, we need  (2+1)-flavor lattice QCD simulations with the $\Lambda\Lambda-N\Xi-\Sigma\Sigma$ coupled channel analysis as well as a careful study on the non-locality of the potential.
The extension of the method outlined in this Letter to this direction is in progress~\cite{sasaki2010}. 

\medskip

\begin{acknowledgments}
We thank authors and maintainer of  CPS++\cite{CPS}, a modified version of which is used in this Letter.
We also thank  the CP-PACS and JLQCD Collaborations \cite{CPPACS-JLQCD}
and ILDG/JLDG \cite{JLDG/ILDG} for providing gauge configurations.
Numerical computations of this work have been carried out 
at KEK supercomputer system (BGL) and
at Univ. of Tsukuba supercomputer system (T2K).
This research is supported in part by MEXT
Grant-in-Aid for Scientific Research on Innovative Areas(No.2004:20105001, 20105003)
and the Large Scale Simulation Program of KEK, Nos.09-23(FY2009) and 09/10-24(FY2010).
S. A. and T. I. are supported in part by MEXT Grant-in-Aid (No.20340047).
N. I. is supported in part by MEXT Grant-in-Aid (No.22540268)
and Grand-in-Aid for Specially Promoted Research (13002001).
T. D. is supported in part by Grant-in-Aid for JSPS Fellows 21$\cdot$5985.
H. N. is supported in part by the MEXT Grant-in-Aid for
Scientific Research on Innovative Areas (No.21105515).
\end{acknowledgments}

%%% REFERENCES %%%

\end{document}